\documentclass[fleqn,twocolumn]{article}

\usepackage[letterpaper,top=2.5cm,bottom=2.4cm,left=2.5cm,right=2.5cm,marginparwidth=1.75cm]{geometry}

\usepackage[english]{babel}
\usepackage[utf8]{inputenc}
\usepackage[T1]{fontenc}

\usepackage{amsmath}%\setlength{\mathindent}{10pt}
\usepackage{siunitx}
\usepackage{bm}
\usepackage{hyperref}

\usepackage{tikz}
\usetikzlibrary{shapes,arrows}
\usepackage{subcaption}

\newcommand{\e}{\mathrm{e}}

\renewcommand{\i}{\mathrm{i}}

\newcommand{\revised}[1]{\textcolor{black}{#1}}

\title{Perceptually Transparent Binaural Auralization of Simulated Sound Fields%\thanks{To whom correspondence should be addressed, e-mail: vesa.valimaki@aes.org. Last updated: June 28, 2021}%
}

\author{Jens Ahrens, \texttt{jens.ahrens@chalmers.se} \\
        Leon Müller, \texttt{leon.muller@chalmers.se}}

\begin{document}

\maketitle

%Abstract
\abstract{% 200 words
Contrary to geometric acoustics-based simulations where the spatial information is available in a tangible form, it is not straightforward to auralize wave-based simulations. A variety of methods have been proposed that compute the ear signals of a virtual listener with known head-related transfer functions from sampling either the sound pressure or the particle velocity (or both) of the simulated sound field. \revised{This article summarizes the most common binaural auralization methods with and without intermediate ambisonic representation of volumetrically sampled sound pressure or sound pressure and particle velocity sampled on spherical or cubical surfaces and presents a perceptual validation thereof. A triangular test ($N=19$) confirmed that all evaluated grids resulted in a perceptually transparent auralization for the three tested sound incidence angles under reverberant conditions. Under anechoic conditions, only the high-density spherical and cubical surface grids lead to  transparent auralization.} All tested methods are available open source in the Chalmers Auralization Toolbox that accompanies this article.% 155 
}

% -----------------------------------------------------------------------------------
% -----------------------------------------------------------------------------------
% -----------------------------------------------------------------------------------
\section{INTRODUCTION}

Wave-based methods like Finite Element Method (FEM), Boundary Element Method (BEM), and Finite-Difference Time-Domain (FDTD) are becoming increasingly popular for room acoustic simulations, and simulations over the entire audible frequency range are possible. Contrary to geometric acoustics-based simulations where the spatial information is available in a tangible form in terms of, for example, incidence directions and spectral fingerprints of reflections~\cite{Vorlander:Springer2008,Barteld:JASA2016,Brinkmann:JASA2019,Blau:AA2021}, it is not straightforward to auralize wave-based simulations, and a variety of approaches have been proposed. 

The output of wave-based simulations may be termed a \emph{virtual sound field}, and the format that the output of any wave-based simulation method can be converted to is the sound pressure and/or particle velocity at a set of observation points, which we term \emph{sampling points} or \emph{sampling nodes}. Those auralization approaches that require sampling of a virtual sound field on an infinitesimal portion of space around the notional receiver point to be auralised may be termed \emph{local}. Examples are~\cite{Southern:TASLP2012}, which uses virtual microphones for obtaining a spherical harmonic (SH) representation of the sampled sound field, and~\cite{Mehra:2014,BilbaoPolitis:2019}, which obtain SH representations via higher-order spatial derivatives. Local approaches require higher-order time integration, which exhibits the inconvenience that it produces a DC drift that is difficult to tame~\cite{HendersonPolitisBilbao:FA2020}.

Non-local approaches include~\cite{Hargreaves:JASA2019}, which obtains SH coefficients by numerically evaluating a boundary integral of pressure and velocity, and~\cite{Stofringsdal:AES2006,Poletti:JASA2008,Sheaffer:TASLP2015,Gomez:MDPI2017}, which sample a portion of space volumetrically. \revised{A comprehensive evaluation of~\cite{Hargreaves:JASA2019} is not available so that we limit our considerations to methods like~\cite{Stofringsdal:AES2006,Poletti:JASA2008,Sheaffer:TASLP2015} that work based sampling the virtual sound field in a portion of space of moderate size.}

Some of the methods mentioned above work solely based on the sampled sound pressure of the simulated field, which can lead to ill-conditioning of the problem. The methods that we consider here are closely related to binaural rendering of microphone array recordings in the literature where this ill-conditioning is well known~\cite{Hulsebos:JAES2002,BalmagesRafaely2007}. The set of sampling nodes in the present case can be interpreted as a virtual microphone array. The main difference is that, contrary to real microphones, the sampling nodes are ideal and their number and placement are virtually unrestricted. Combining pressure and pressure gradient can greatly improve the conditioning of the problem~\cite{BurtonMiller:RSL2971,Hulsebos:JAES2002}. One incarnation of this combination are (virtual or actual) microphones with cardioid directivity~\cite{BalmagesRafaely2007,Thomas:ICASSP2019}, which are a promising solution to the present problem~\cite{Cosnefroy:I3DA2023}.

We want to emphasize that our work does not evaluate acoustical simulations themselves. Our purpose is to investigate how non-geometrical acoustics simulations can be auralized binaurally, i.e.~how to most favorably compute signals from the output of the simulation that would arise at a listener's ears if the listener were exposed to the simulated sound field. As it is likely that it will not be possible to achieve that the output of the auralization is numerically identical to the ground truth, we seek for \emph{perceptually transparent} auralization, i.e.~auralization that is perceptually indistinguishable from the ground truth in a direct comparison. 

Our investigation is not based on the output of an actual simulation framework. We rather use a representation of room acoustic responses from which we can compute both a binaural ground truth signal as well as sampled sound field data the auralization of which we evaluate against the binaural ground truth. Previous perceptual evaluations of auralization of non-geometric acoustics simulations used auralization of limited scope or frequency range (or both)~\cite{Gomez:MDPI2017,Pathre:EAA2022,Yoshida:MDPI2023}. The potential for general perceptual transparency is therefore unclear.

The auralization solution that we seek in this manuscript is intended to be as general as possible, which means that we only consider solutions that 1) do not make assumptions on the sound field that is to be auralized, and that 2) can work on running signals. The motivation for 1) is that the auralization is desired to be applicable to any simulated sound field. There are methods that were originally proposed in the field of auralization of microphone array measurements of room impulse responses such as~\cite{Merimaa:JAES2005,Gunnarsson:TASLP2021} that parameterize the sound field, for example into direct and diffuse components. This can indeed provide a benefit over non-parametric methods, especially when the number of sampling nodes is small. Yet, there is uncertainty in the output from parametric methods if non-standard sound fields are considered such as sound fields outdoors, in small compartments like in car cabins where it is not straightforward to differentiate direct sound from early reflections, or if occlusion of the direct sound occurs. 

Linear and time-invariant (LTI) sound field simulations can be expressed in terms of impulse responses. While this is likely to be the most common application area of auralization today, we still formulate requirement 2) so that also dynamic scenarios and scenarios that comprise nonlinearities such simulations of electroacoustic systems can be handled.
Parametric rendering methods have also been proposed for running signals~\cite{Pulkki:JAES2007,Politis:ICASSP2018}. We exclude those from the present investigation for the reasons stated above. 

For convenience, our investigation is performed based on simulated room impulse responses. This does not limit the generalizability of the results as all methods can be straightforwardly applied to running signals without modification.

% -----------------------------------------------------------------------------------
% -----------------------------------------------------------------------------------
% -----------------------------------------------------------------------------------
\section{Auralization Framework}\label{sec:methods}

What is common to all auralization methods that we consider here is that the auralization itself is modelled as an LTI system and is represented by a set of impulse responses. The set of impulse responses essentially represents the multiple-input and multiple-output (MIMO) transfer path from the sampling nodes to the ears of the listener. Refer to Fig.~\ref{fig:flow_chart} for an illustration. The transfer path therefore comprises the head-related transfer function (HRTFs) of the listener as well as their head orientation. Once the impulse responses have been computed, carrying out the auralization consists in performing the required convolutions. This process if often termed \emph{rendering} in spatial audio. The result of the rendering is the ear signals that would arise at the listener's ears if they were exposed to the virtual sound field with a given head orientation. 

\tikzstyle{block} = [draw,fill=gray!20,minimum size=2em]
\tikzstyle{circle} = [draw,shape=circle,fill=gray!20,minimum size=2em]
\begin{figure}[tbh]
\begin{center}

\begin{tikzpicture}[>=latex']

    \node at (0, 5.2) (input1) {$s(\vec{\mathbf{x}}, t)$};
    
    \node[block] at (0, 3.25) (block2) {Binaural rendering};
    \draw[->] (input1) -- node[strike out,draw,-]{} (block2);

    \node[] at (.27, 4.25) {\scriptsize ~~~343};
    \node[] at (.27, 2.25) {\scriptsize 2};
    
    \node at (0, 1.25) (output) {$b(t)$};

    \draw[->] (block2) -- node[strike out,draw,-]{} (output);
    
\end{tikzpicture}
\hspace{1.2cm}
\begin{tikzpicture}[>=latex']

    %\node at (0, 5.2) (input1) {$s(t)$};
    \node at (0, 5.2) (input1) {$s(\vec{\mathbf{x}}, t)$};
    \node[block] at (0, 4) (block1) {SH decomposition};
    \draw[->] (input1) -- node[strike out,draw,-]{} (block1);

    \node[] at (.27, 4.65) {\scriptsize ~~~343};
    
    \node[block] at (0, 2.5) (block2) {Binaural rendering};
    \draw[->] (block1) -- node[strike out,draw,-]{} (block2);

    \node[] at (.27, 3.25) {\scriptsize 81};
    \node[] at (.27, 1.85) {\scriptsize 2};
    
    \node at (0, 1.25) (output) {$b(t)$};

    \draw[->] (block2) -- node[strike out,draw,-]{} (output);
    
\end{tikzpicture}
\caption{Block diagram of the signal processing pipeline for direct auralization (left) and ambisonic auralization (right). $s(\vec{\mathbf{x}}, t)$ are signals at the sampling nodes $\vec{\mathbf{x}}$ and $b(t)$ is the binaural output. Each of the gray blocks represents a MIMO convolution. In these examples, 343 sampling nodes are assumed as well as an ambisonic representation of order $N=8$, which produces an ambisonic signal with~$81$ channels. In either case, the binaural rendering stage takes the instantaneous head orientation of the listener into account.}
\label{fig:flow_chart}
\end{center}
\end{figure}

The available methods may be divided in two categories: Methods that we term here \emph{direct methods} obtain the binaural output signals directly from the signals at the sampling nodes. Methods that we term here \emph{ambisonic methods} produce a representation of the sampled sound field in terms of spherical harmonic basis functions (SHs). This representation is also referred to as ambisonic representation and can be rendered binaurally or using a loudspeaker array~\cite{Zotter:book2019}. The advantage of the ambisonic methods is that the problem of auralization is decomposed into two independent stages that can be optimized separately. From a purely practical perspective, head-tracked rendering of the ambisonic representation can be carried out straightforwardly by applying rotation operations on either the ambisonic sound field or the ambisonic HRTF representation. The application of head tracking on the rendering side in direct auralization requires computing the auralization separately for each possible head orientation of the listener, which can be a substantial computational burden. 

\begin{figure}[tb]
    \centering
    %                         left bottom right top
    \includegraphics[width=\columnwidth,trim=0in 0in 0in 0in, clip=true]{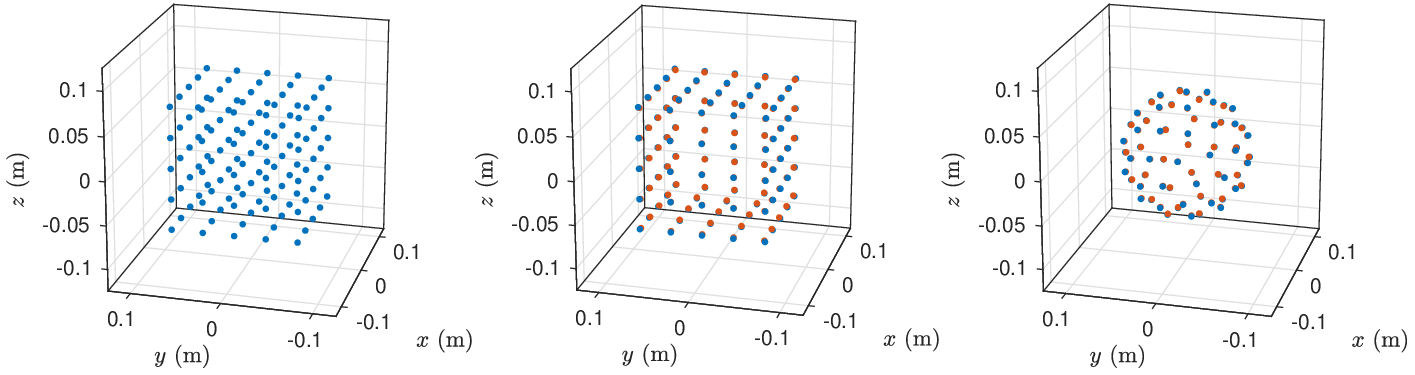}
    \caption{The employed sampling grids. Left: Volumetric cube (125 nodes). Middle: Cubical surface grid (98 double nodes). Right: Spherical surface grid (64 double nodes). With the surface grids, blue color indicates a pressure node, and orange color indicates a coincident velocity node. The diameter of the spherical grid and the edge length of the cubical grids are~\SI{140}{mm}.}
    \label{fig:grids}
    %\vspace{-2mm}
\end{figure}

The following subsections outline the signal processing in the auralization methods that we investigated. For the purpose of reproducibility, we provide all our implementations in the Chalmers Auralization Toolbox~\cite{Ahrens:AA2023,Ahrens:auralization_toolbox}. As we will discuss below, we extended the available methods such that both ambisonic auralization and direct auralization can be carried out based on cubical volumetric sampling (CV) and based on spherical and cubical surface sampling (SS and CS) of the virtual sound field. Fig.~\ref{fig:grids} depicts examples for all three types of sampling grids. \revised{All elaborated further below, it is sufficient to sample only the sound pressure if a volumetric grid is used. Surface grid require sampling both sound pressure and particle velocity.}

While all investigated methods are formulated in frequency domain, it appears more flexible if the auralization is carried out in practice via block-wise processing of time-domain signals because the auralization does then not need to be adapted to the length of the room response. This is also how the Chalmers Auralization Toolbox performs it. We highlight here that this presents challenges regarding the implementation for a variety of reasons. For example, the DC part of a given spectral representation can be undefined, or time aliasing can occur, which are both challenges that do not occur in pure frequency-domain considerations. The discussion of the implementation details is beyond the scope of this article, and we refer the reader to the documentation and the code of the Chalmers Auralization Toolbox.

%

% -----------------------------------------------------------------------------------
\subsection{The Ambisonic Methods}

The ambisonic methods that we investigate are based on~\cite{Sheaffer:TASLP2015}, which is formulated for volumetrically sampled sound pressure fields and may be considered an extension of~\cite{Stofringsdal:AES2006}. The SH representation is obtained via inversion of a matrix of the SH basis terms. Contrary to~\cite{Sheaffer:TASLP2015}, we use a 
singular value decomposition (SVD)~\cite{Cohen:svdbook} that is regularized by limiting the dynamic range of the singular values to regularize the matrix inversion instead of soft-clipping of the matrix elements followed by an unregularized pseudo inverse. \revised{We will discuss the significance of the choice of regularization in Sec.~\ref{sec:discussion}}. 

Volumetric sampling may be inefficient in that it may require unnecessarily many sampling points to yield a well-conditioned problem. The Kirchhoff-Helmholtz integral~\cite{Gumerov:book2005} demonstrates that the sound pressure field inside a source-free domain is uniquely described by the sound pressure $S(\vec{\mathbf{x}}, \omega)$ and normal sound pressure gradient $\frac{\partial}{\partial \Vec{\mathbf{n}}} S(\Vec{\mathbf{x}}, \omega)$ distributions along the simply connected surface that encloses the domain. It is proven in~\cite[p.~207]{BurtonMiller:RSL2971} that a weighted sum $S(\vec{\mathbf{x}}, \omega) + \gamma \frac{\partial}{\partial \Vec{\mathbf{n}}} S(\Vec{\mathbf{x}}, \omega)$ uniquely defines the sound pressure inside the volume that is enclosed by the surface so long as $\mathrm{Im}\{\gamma\}\neq 0$. Choosing $\gamma =  1 / (\i \frac{\omega}{c})$ fulfills the uniqueness criterion and is a particularly convenient choice as it assures that pressure and gradient are added with similar magnitudes. It also allows for the weighted sum to be interpreted as a virtual sensor with (far-field) cardioid directivity, which connects our formulation well to the literature on microphone arrays~\cite{BalmagesRafaely2007,Thomas:ICASSP2019}. This enables the employment of surface sampling grids, which may potentially be more efficient than volumetric ones. 

The sound pressure gradient can be straightforwardly computed from the particle velocity~\cite{Williams:book1999}, which a number simulation methods provide straightforwardly. If the particle velocity is not available, then it can be obtained from sampling the sound pressure along two surfaces that are much closer to each other than the wavelength. %We confirmed through informal listening that the binaural output signals produced from a double pressure layer are perceptually identical to the binaural output signals produced from pressure and velocity along one surface layer so long as the distance between layers is not larger than \SI{5}{mm}.

Extending~\cite{Sheaffer:TASLP2015} to spherical surface grids with radially outward-facing cardioid sensors is straightforward given the literature on spherical microphone arrays~\cite{BalmagesRafaely2007}. The mathematical details are provided in the documentation of the Chalmers Auralization Toolbox. As to our awareness, cardioid sensors have only been used with circular and spherical contours for performing SH sound field decomposition. This is unfortunate as some wave-based simulation methods such as FDTD can employ Cartesian sampling schemes that would require interpolation to realize a spherical sampling grid. To bridge this gap, we extended the method from~\cite{Sheaffer:TASLP2015} to arbitrary simply connected surface grids with normally outward-facing cardioid sensors. We included cubical surface grids into the present investigation. This solution was not available in the literature and is an original contribution by the authors. The mathematics behind it is not straightforward, and we refer the reader to the documentation of the Chalmers Auralization Toolbox for details~\cite{Ahrens:CAT2025}. 

There are two main sources for systematic inaccuracies in ambisonic auralization: 1) Spatial aliasing due to the discretization of the sampled sound field and 2) truncation of the order $N$ of the SH representation of the sound field (and consequently of the HRTFs). Spatial aliasing typically increases the energy at high frequencies, and order truncation decreases or increases it depending on the incidence direction~\cite{Ahrens:DAGA2024}. Anticipating the interaction between the two is not straightforward. 

\begin{figure}[tb]
    \centering
    %                         left bottom right top
    \hfill\includegraphics[height=.42\columnwidth,trim=0in 0in 0in 0in, clip=true]{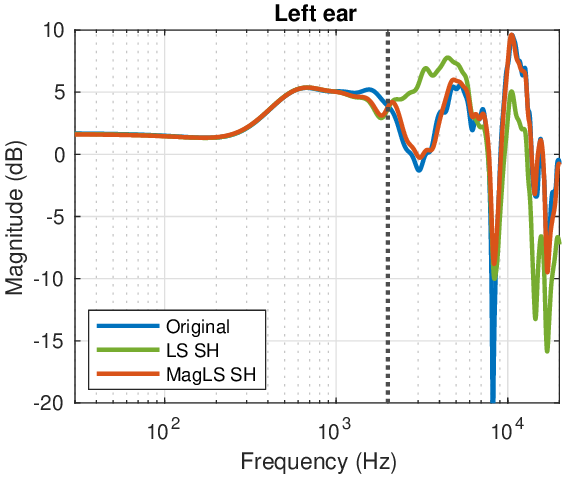}\hfill\includegraphics[height=.42\columnwidth,trim=0in 0in 0in 0in, clip=true]{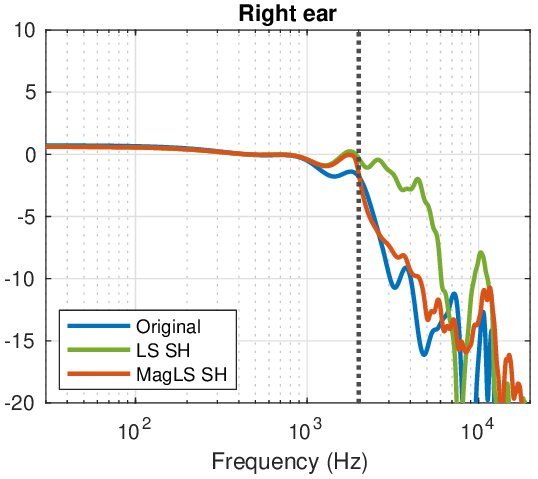}
    \caption{The employed HRTFs (blue line) as well as their SH representations for sound incidence from the left computed from a 4th-order SH decomposition using the conventional least-squares (LS) fit of SH coefficients (green line) as well as using a MagLS fit (orange line). The gray dotted line indicates the frequency below which LS and above which MagLS was used to compute the SH coefficients in the case that is labeled MagLS.}
    \label{fig:magls}
    %\vspace{-2mm}
\end{figure}

Contrary to the original formulations of ambisonic auralization~\cite{Stofringsdal:AES2006,Sheaffer:TASLP2015}, we employ a magnitude least-squares (MagLS) SH representation of the HRTFs~\cite{Schorkhuber:DAGA2018}, which essentially eliminates the effect of the order truncation on the HRTFs. MagLS SH representations achieve an increase of the accuracy of the magnitude at the expense of a modification of the phase. It was demonstrated in~\cite{Schorkhuber:DAGA2018} that the phase modification is not audible if it occurs at frequencies higher than \SI{3}{kHz}. This condition is fulfilled for all our data that use an SH order of~6 or higher.
Fig.~\ref{fig:magls} illustrates the increase of the accuracy of the magnitude exemplarily for an SH order of $N=4$. \revised{Fig.~\ref{fig:data} (left column) depicts numerical data for example auralizations.}

The effects of spatial aliasing and order limitation in the decomposed sound field occur in the same frequency range. Their effect on the distribution of the energy over the frequency axis is hardly dependent on the incidence direction of the sound field for the grids that we consider here. We therefore equalize the sound field decomposition by computing a global minimum-phase filter that minimizes the deviation of the binaural output from the ground truth averaged over many sound incidence directions. This equalization filter is only effective in the frequency range where spatial aliasing is apparent. Recalling that HRTFs are defined as the acoustic ear signals due to a plane wave in free-field conditions makes sampled plane waves and HRTFs a convenient set of input/output data based on which this equalization filter can be computed.

Independent of the sampling grid, the ambisonic methods produce an SH representation of the sampled sound field. The binaural rendering of these is well established and detailed, for example, in~\cite{Ahrens:binaural_rendering_in_sh}.

%

% -----------------------------------------------------------------------------------
\subsection{The Direct Methods}

The direct auralization methods are based on~\cite{Poletti:JASA2008}. The auralization is modeled as a MIMO system. The input to the MIMO system is the sampled sound field data. The output are the left and right ear signals. The set of filters that represent the transfer path between the input and the output are computed through a least-squares (LS) fit on sample data, which is regularized by limiting the dynamic range of the singular values of an SVD. Sampled plane waves and HRTFs are a convenient set of input/output data for computing the LS solution. The original method was demonstrated for volumetrically sampled sound pressure. Applying it to spherical and cubical surface sampled data does not require any modifications. 

\begin{figure*}[htb!]
    \centering
    \begin{subfigure}{\columnwidth}
        %                         left bottom right to
        \includegraphics[height=.65\columnwidth,trim=0in 0in 4.8in 0in, clip=true]{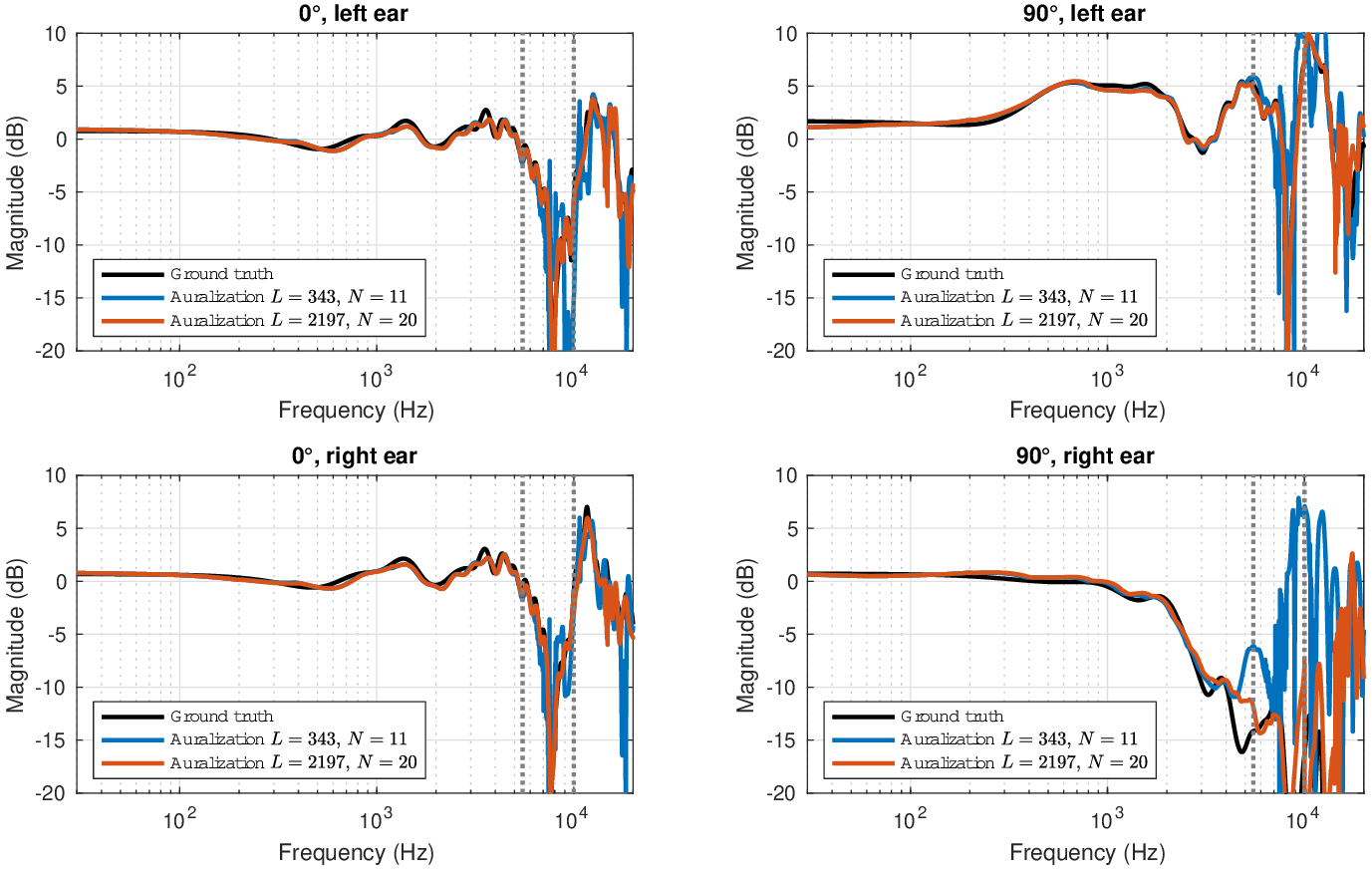}\includegraphics[height=.65\columnwidth,trim=5.1in 0in 0in 0in, clip=true]{brtfs_ambisonics_cubical_volume.eps}\hfill
        \caption{Cubical volume grid, ambisonic auralization}
        \label{fig:data1}
    \end{subfigure}
    \begin{subfigure}{\columnwidth}
        %                         left bottom right to
        \hfill\includegraphics[height=.65\columnwidth,trim=0in 0in 4.8in 0in, clip=true]{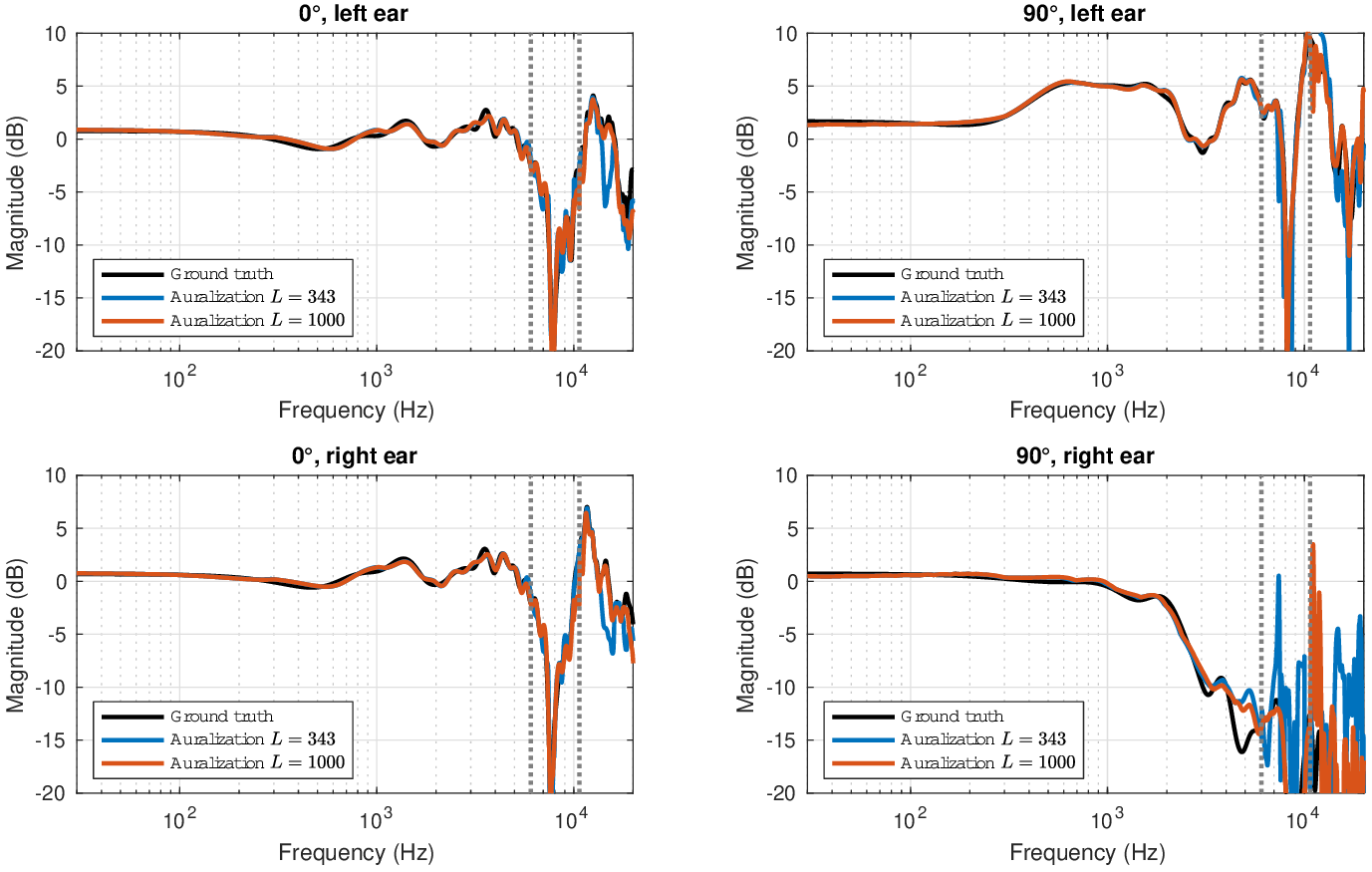}\includegraphics[height=.65\columnwidth,trim=5.1in 0in 0in 0in, clip=true]{brtfs_direct_cubical_volume.eps}
        \caption{Cubical volume grid, direct auralization}
        \label{fig:data2}
    \end{subfigure}
    \begin{subfigure}{\columnwidth}
        %                         left bottom right to
        \includegraphics[height=.65\columnwidth,trim=0in 0in 4.8in 0in, clip=true]{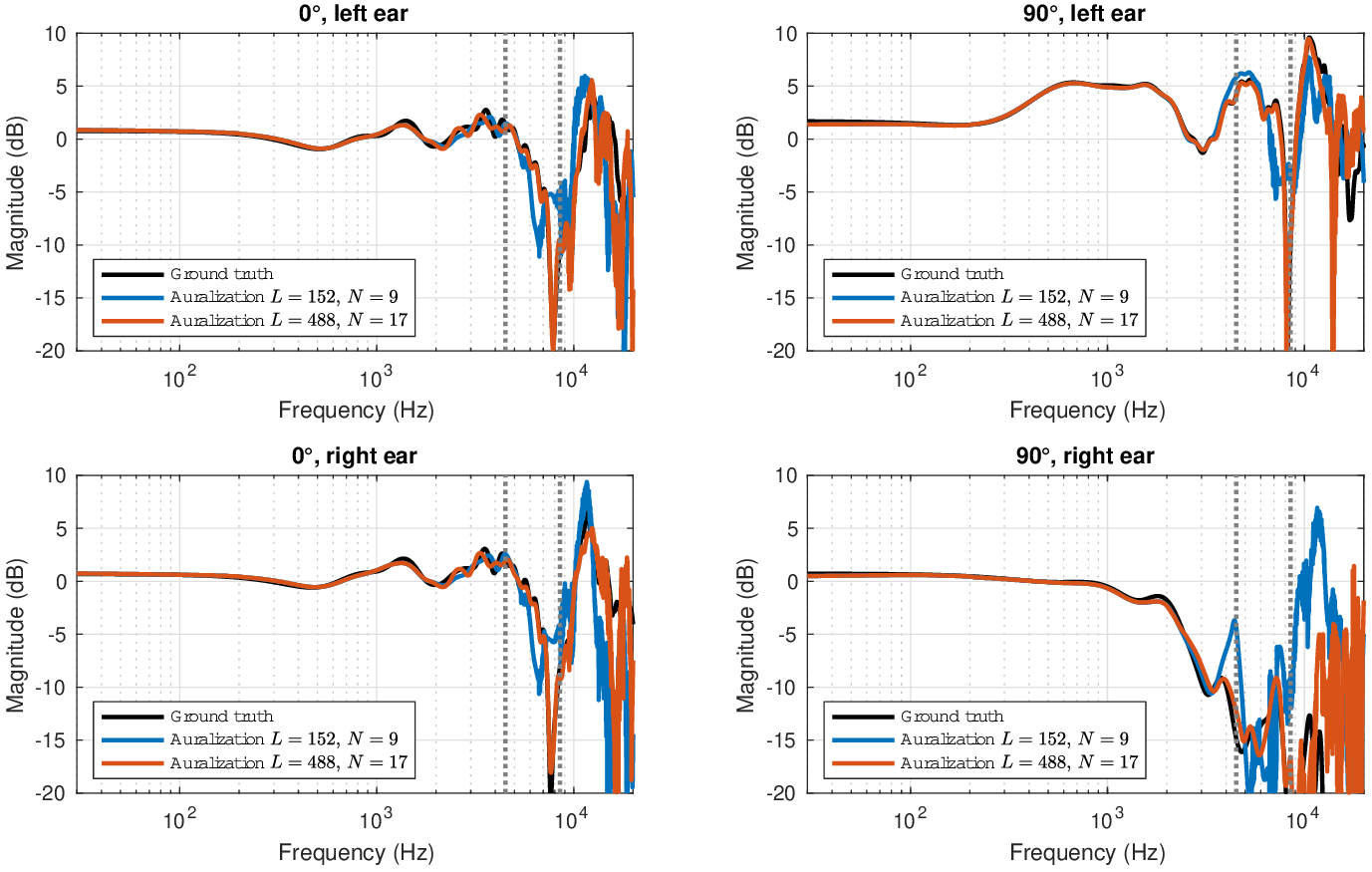}\includegraphics[height=.65\columnwidth,trim=5.1in 0in 0in 0in, clip=true]{brtfs_ambisonics_cubical_surface.eps}\hfill
        \caption{Cubical surface grid, ambisonic auralization}
        \label{fig:data5}
    \end{subfigure}
    \begin{subfigure}{\columnwidth}
        %                         left bottom right to
        \hfill\includegraphics[height=.65\columnwidth,trim=0in 0in 4.8in 0in, clip=true]{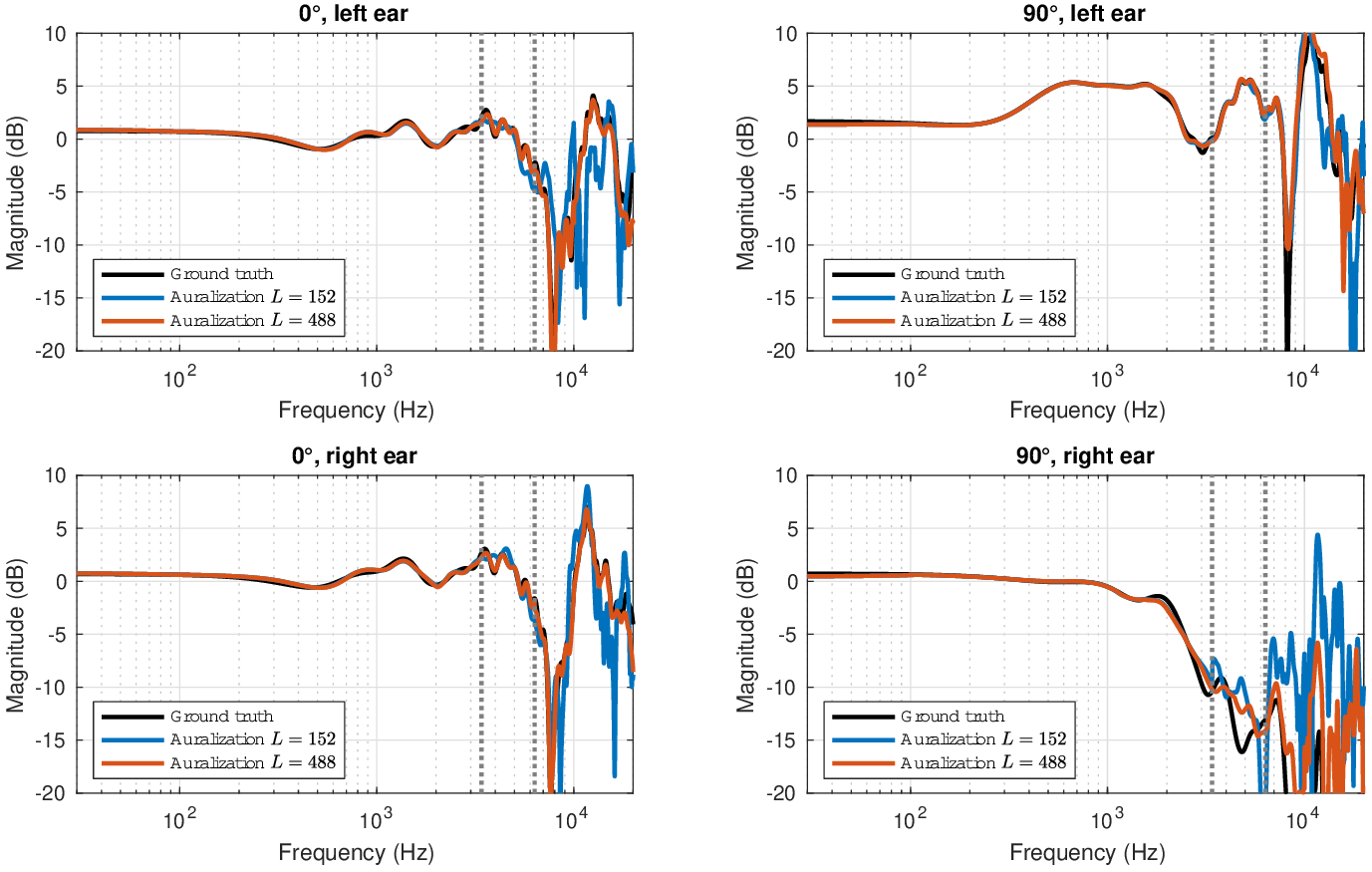}\includegraphics[height=.65\columnwidth,trim=5.1in 0in 0in 0in, clip=true]{brtfs_direct_cubical_surface.eps}
        \caption{Cubical surface grid, direct auralization}
        \label{fig:data6}
    \end{subfigure}
    \begin{subfigure}{\columnwidth}
        %                         left bottom right to
        \includegraphics[height=.65\columnwidth,trim=0in 0in 4.8in 0in, clip=true]{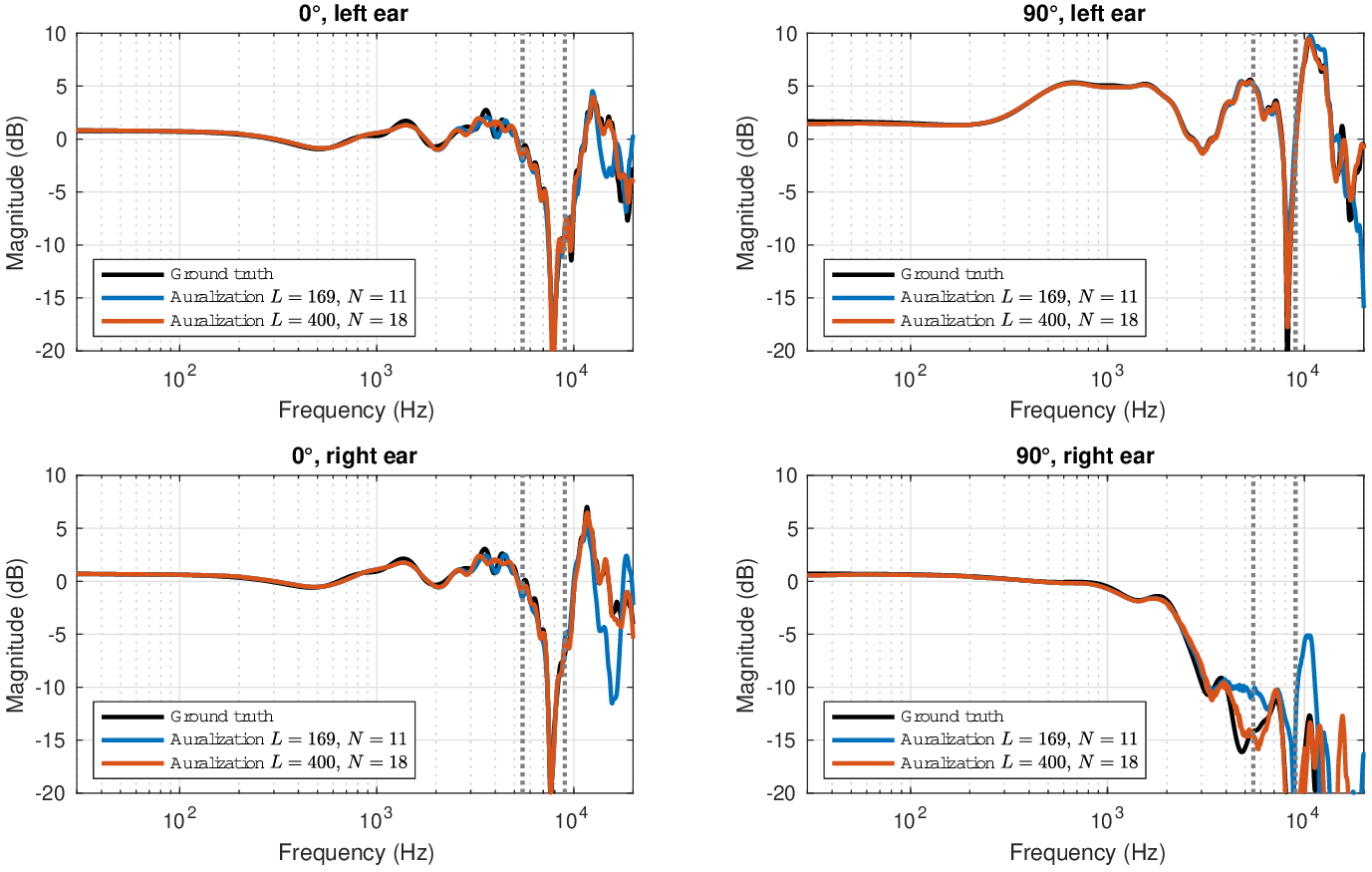}\includegraphics[height=.65\columnwidth,trim=5.1in 0in 0in 0in, clip=true]{brtfs_ambisonics_spherical_surface.eps}\hfill
        \caption{Spherical surface grid, ambisonic auralization}
        \label{fig:data3}
    \end{subfigure}
    \begin{subfigure}{\columnwidth}
        %                         left bottom right to
        \hfill\includegraphics[height=.65\columnwidth,trim=0in 0in 4.8in 0in, clip=true]{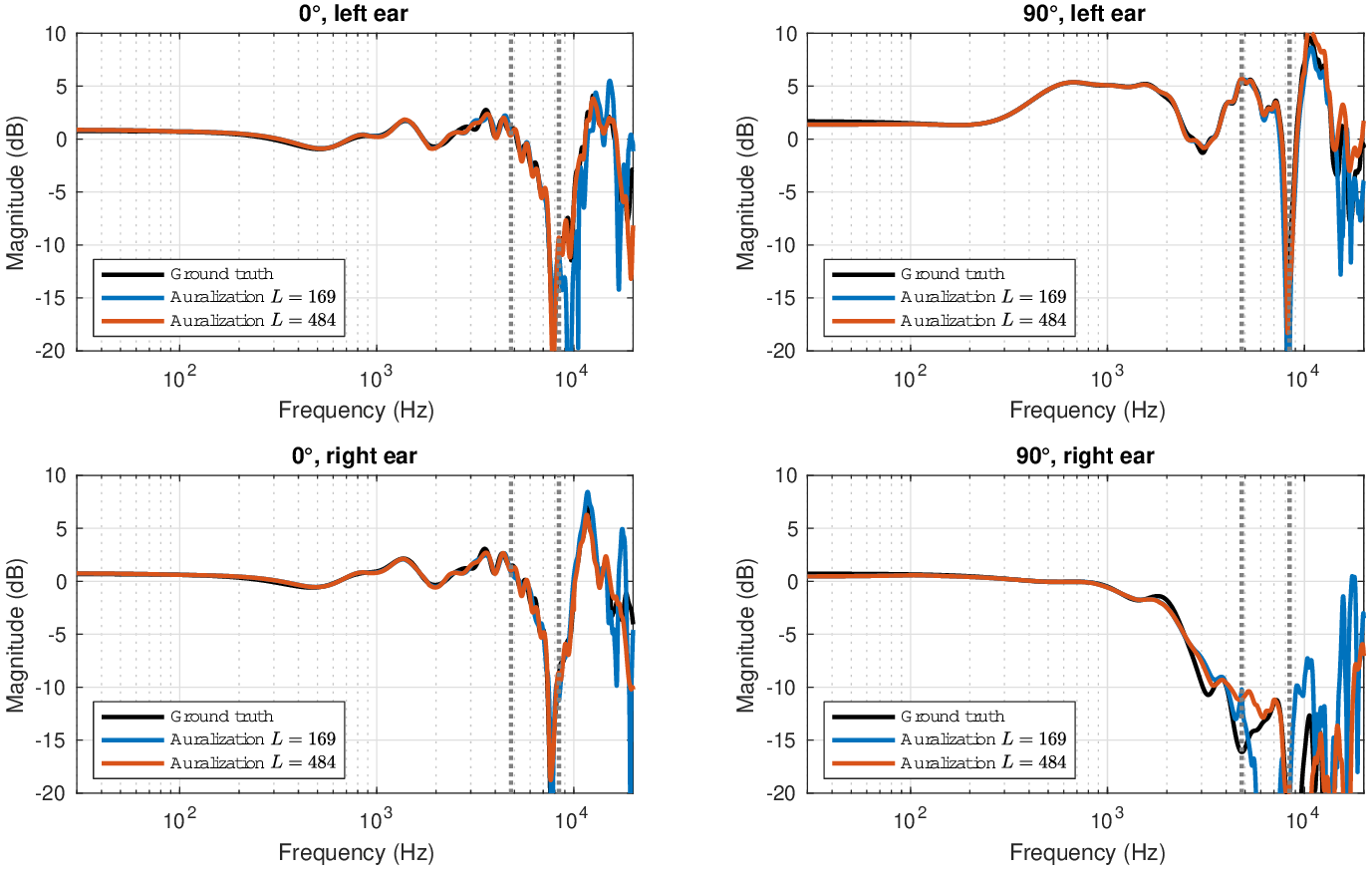}\includegraphics[height=.65\columnwidth,trim=5.1in 0in 0in 0in, clip=true]{brtfs_direct_spherical_surface.eps}
        \caption{Spherical surface grid, direct auralization}
        \label{fig:data4}
    \end{subfigure}
    \caption{\revised{Example anechoic binaural data for each combination of grid and method for a grid with a moderate number of sampling points and a high-density grid each (Left column: ambisonic auralization. Right column: direct auralization). The binaural output is due to a plane wave with incidence from straight ahead or from \SI{90}{\degree} to the left. The ground truth is the HRTF corresponding to the given incidence direction. The vertical dotted lines indicate above what frequency MagLS-HRTFs are employed (ambisonic methods) or the eMagLS solution is employed (direct methods). All data were computed by performing the auralization on the time-domain signals.}}
    \label{fig:data}
    %\vspace{-2mm}
\end{figure*}

Like any other method, also direct auralization suffers from spatial aliasing \revised{(cf.~the numerical data depicted in Fig.~\ref{fig:data} (right column))}. The LS solution can exhibit large errors in the frequency range where spatial aliasing occurs. We found that the solution is perceptually more favorable if the LS solution is replaced with a MagLS solution in the frequency range where the spatial aliasing occurs. Our implementation corresponds to what is termed end-to-end MagLS, variant 2, in~\cite{Deppisch:I3DA2021}. 

% -----------------------------------------------------------------------------------
% -----------------------------------------------------------------------------------
% -----------------------------------------------------------------------------------
\section{Grid Parameters}

Choosing the parameters for a grid with a given number of nodes $L$ is a compromise: A grid with large dimensions is desired for being able to resolve spatial information at low frequencies. Keeping the spatial aliasing frequency high requires a small spacing between the nodes. It was demonstrated in~\cite{Ahrens:DAGA2024} that high SH order spatial information at low frequencies in the sound field does not reach the ear of the listener because it is suppressed by the directivity of the ears. The sampling grid can therefore be small compared to the wavelength at those low frequencies where detailed spatial information is not required. This produces a high spatial aliasing frequency, which is favorable and comes at virtually no cost. We verified using the approach from~\cite{Ahrens:DAGA2024} that a diameter or edge length of \SI{0.14}{m} is the smallest size that avoids compromises in the binaural output signals due to the required regularization. We chose it for all grids. We could not find a noteworthy benefit in choosing a different grid size for any of the investigated configurations.

Solutions for optimal placement of sampling points in a volume or on a surface contour for sound field decomposition have been proposed in the literature~\cite{Chardon:IEEE2015,Koyama:IEEE2020,Verburg:JSM2024}. These have only been demonstrated in narrow frequency ranges. It is unclear how they can be extended to the entire audible frequency range that is of interest here. We therefore only consider evenly sampled grids in this article. This means that only certain numbers of sampling nodes are possible for cubical volumetric and cubical surface grids. We selected the layouts for spherical surface grids from Fliege grids~\cite{Fliege:JNA1999} with comparable numbers of nodes.

For ambisonic auralization, the frequency above which spatial aliasing occurs can be estimated with the $N=kR$-rule where $N$ is the maximum order that the array provides, $k=2\pi f_\text{a} /c$ is the wavenumber that is considered, $R$ is the radius of the baffle, and $c$ is the speed of sound~\cite{Kennedy:IEEE2007}. Solving the expression for the aliasing frequency $f_\text{a}$ yields
\begin{equation}\label{eq:N=kr}
    f_\text{a} = \frac{N}{2\pi R} c  \, .
\end{equation}
The maximum order $N$ that Fliege grids with $L$ nodes on spherical surfaces provide is given by $(N+2)^2 = L$. As an example, a spherical surface grid with $R=\SI{0.07}{m}$  and $L = 144$ supports $N=10$ and produces an aliasing frequency of $f_\text{a} = \SI{7.8}{kHz}$. Slightly more nodes appear to be required for cubical surfaces for a given $N$ and quite a few more for volumetric grids. The aliasing frequency $f_\text{a}$ does not seem to differ substantially between ambisonic and direct methods.

%

% -----------------------------------------------------------------------------------
% -----------------------------------------------------------------------------------
% -----------------------------------------------------------------------------------
\section{Computation of the Ground Truth}

Validation of the perceptual transparency of the auralization requires comparing the auralization to the ground truth ear signals. Anechoic conditions are straightforward given that HRTFs are defined as the acoustic ear signals due to a plane wave in free-field conditions. We therefore used spatially sampled plane waves as the input to the auralization and the HRTFs as the ground truth in this setting. The sound pressure $S(\omega, \vec{x})$ at radian frequency $\omega$ and at location $\vec{x}$ due to a plane wave is given by $S(\omega, \vec{x}) = \e^{-i \vec{k}_\text{pw}^T \vec{x}}$ where $\vec{k}_\text{pw}^T$ is the transposed wave vector of the plane wave. The sound pressure gradient is computed accordingly.

Reverberant conditions are less straightforward in that no wave-based simulation method is free of systematic errors, and it is not straightforward how to compute the ground truth ear signals. Simpler room acoustic simulation methods like the image source method allow for computing a ground truth, but the resulting reverberation can sound unnatural. 

\begin{figure}[tb]
    \centering
    %                         left bottom right top
    \hspace{8mm}\includegraphics[width=.3\columnwidth,trim=0in 0in 0in 0in, clip=true]{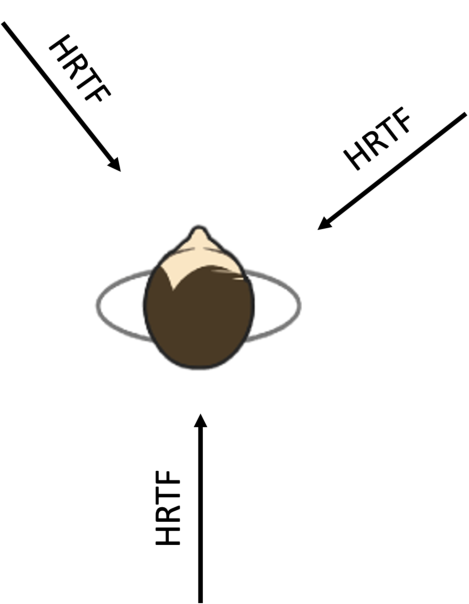}\hfill\includegraphics[width=.5\columnwidth,trim=0in 1.3in 0in 0in, clip=true]{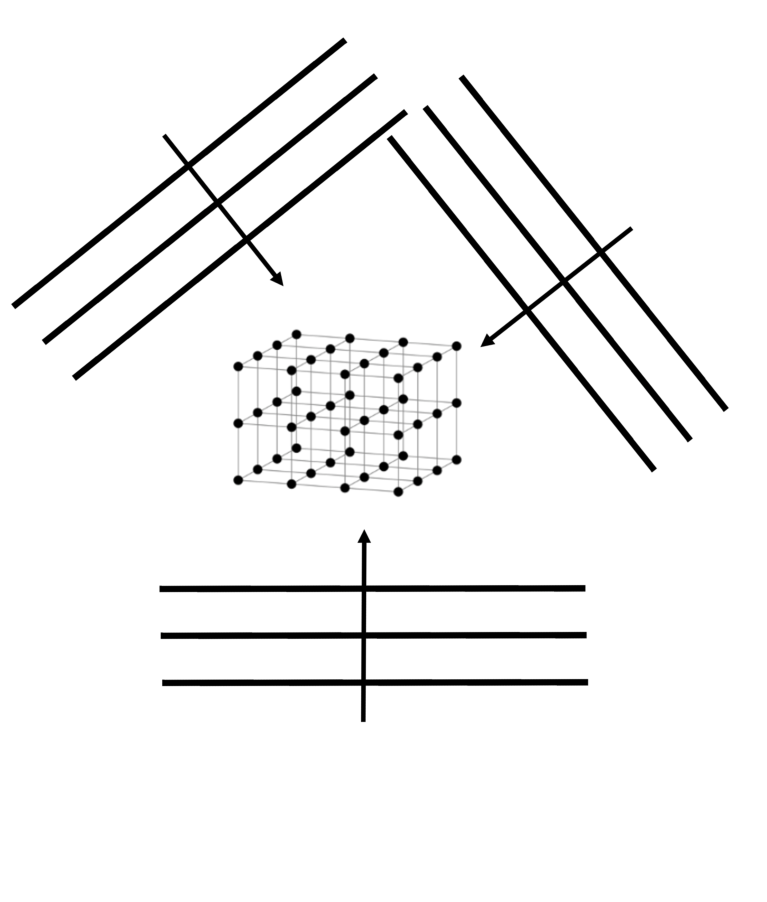}
    \caption{Illustration of the computation of the ground truth binaural signal (left) and the sampled sound field data (right) based on SDM data as performed for the reverberant condition.}
    \label{fig:sdm}
    %\vspace{-2mm}
\end{figure}

We therefore chose the following procedure: The spatial decomposition method (SDM)~\cite{Tervo:JAES2013} produces a representation of a room impulse response in terms of the pressure impulse response as well as an incidence direction for each digital sample of the pressure impulse response. A binaural representation of the room impulse response can be obtained by assigning the HRTF that corresponds to the incidence direction of a given digital sample to that digital sample and superposing the weighted HRTFs for all digital samples. This binaural representation has been shown to be perceptually slightly different from the actual binaural room response. We chose it for the present purpose because it was confirmed by different authors that the perceptual result is highly plausible~\cite{Kaplanis:PhD2017,Ahrens:WASPAA2019}. The input data to the auralization can be computed from the SDM data by assigning plane waves with appropriate incidence directions instead of HRTFs to each digital sample. Fig.~\ref{fig:sdm} illustrates the concept. An implementation is available at~\cite{Ahrens:acoustic_room_responses}.

%

% -----------------------------------------------------------------------------------
% -----------------------------------------------------------------------------------
% -----------------------------------------------------------------------------------
\section{Perceptual Experiment}

%\color{blue}

The experiment was a triangular test to determine if a difference between the stimulus and the ground truth was detectable~\cite{Maynard:book1965, ISO4120:2021, Zacharov2018}. The subjects were presented with a graphical interface with three buttons in each trial. Two of the buttons played identical signals (ground truth or auralization), and the third one played the corresponding other signal. The assignment of the buttons as well as the order of the conditions were randomized for each participant. The subject's task was to report via a button click which of the three stimuli sounded different from the other two. They were instructed to click any of the three answer buttons randomly if they were not detecting an audible difference. The subjects could listen to the stimuli as many times as they wanted. Switching from one stimulus to another started playback of the source signal from the beginning.

We choose a 5-second segment of a rock drum rhythm as the test signal. It is transient, which makes it suitable for investigating the auralization of reverberation, and it is one of the practically relevant signals that has relatively much energy at high frequencies, which ensures that the consequences of spatial aliasing are reflected in the binaural output with sufficient prominence. 

We presented preliminary results from a similar validation study on the Chalmers Auralization Toolbox in~\cite{Ahrens:DAGA2025}. That experiment used a transformed two-up/one-down staircase procedure~\cite{Levitt:JASA1971} to determine the minimum required number of sampling points for making the difference between the stimulus and ground truth to be undetectable. Head tracking was employed, and the subjects were able to switch the principal direction of the sound source between straight ahead and lateral. Further experimentation with the same signals suggested to us that the amount of degrees of freedom that the subjects in the experiment from~\cite{Ahrens:DAGA2025} were required to master was so high that it produced a significant number of unforced errors. Only a limited amount of data was therefore available for grids with high density of sampling points, which made it difficult to interpret the results. Furthermore, the experiment was based on the assumption that reducing the energy of spatial aliasing to a sufficient extent by increasing the number of sampling points will necessarily lead to perceptually transparent auralization. We will discuss in Sec.~\ref{sec:discussion} below why we are doubtful that this assumption is justified.

We chose to present a study here that aims at validating perceptual transparency for a smaller set of auralization parameters in a more robust manner than~\cite{Ahrens:DAGA2025} at the cost of not identifying the minimum required number of sampling points. The present experiment does not employ head tracking, and the directions of incidence of the direct sound are preselected to be 0\textdegree, -45\textdegree, and 90\textdegree. It is important to include a variety of sound incidence directions as it was confirmed in~\cite{Zaunschirm:JASA2018,Ahrens:JASA2019} and other locations in the literature that lateral sound incidence is the most revealing one for detecting perceptual differences between reference and stimulus. This is presumably because the target signal at the contralateral ear exhibits comparatively low energy so that artifacts, for example due to spatial aliasing, become perceptually more prominent. The artifacts at the contralateral ear are even numerically prominent for most conditions depicted in Fig.~\ref{fig:data}.

The room acoustic conditions that we tested were anechoic and a hall with a reverb decay time of \SI{1}{s}. Loudness alignment of the stimuli was not necessary as all tested methods preserve the signal level. The playback level was the same for all subjects and was similar to that of conversational speech.

We chose to include both low-density grids as well as high-density grids in the study. Low-density grids exhibit so few sampling points that it is expected that the subjects will be able to reliably differentiate the auralization from the ground truth. The purpose of including these grids in the study was to ensure that there are conditions for which the subjects are able to carry out their task with great confidence. We omitted including low-density grids for the reverberant condition to minimize the experiment duration. It is not to be expected for these low-density grids that the subjects' responses to reverberant stimuli would be considerably different than to the anechoic stimuli~\cite{Ahrens:DAGA2025}.

We chose the high-density grids such that exhibited so many sampling points that we found a saturation of the perceptual difference between the auralization and the ground truth in pilot studies. In other words, it appeared that a further increase of the number of sampling points in the high-density grids did not reduce the perceptual difference to the ground truth.

Tab.~\ref{tab:grid_sizes} summarizes the conditions that were tested in the experiment. The experiment comprised a total of 54 conditions that were tested (six anechoic conditions with low-density grids, six anechoic conditions with high-density grids, and six reverberant conditions with high-density grids; three different sound incidence directions for each of them) preceded by a training comprising eight conditions.

%lists all investigated grid layouts. We found that it was not useful to use cubical volumetric grids with an SH order of higher than $N=18$. Example binaural data are provided in Fig.~\ref{fig:data}. The different parameters produce primarily differences only at high frequencies where spatial aliasing limits the accuracy.

%
\begin{table}[ht]
\centering
\small
\begin{tabular}{l|l|l|l}
%$L_\text{cub.~vol.}$ & $L_\text{cub.~surf.}$ & $L_\text{sph.~surf.}$ \\ \hline
 & Low density & High density& High density \\ 
 & (anechoic) & (anechoic) & (reverberant)\\\hline
CV(a)  & 216 (7) & 2197 (20) & 2197 (20) \\ 
CV(d) & 27  & 1000  & 1000  \\
CS(a)  & 98 (7) & 488 (17) & 488 (17) \\ 
CS(d) & 98  & 488  & 488  \\
SS(a)  & 25 (3) & 400 (18) & 400 (18)\\ 
SS(d) & 25 & 484  & 484 \\[1ex]
\end{tabular}
\caption{Number $L$ of data points for the different grids for ambisonic (a) and direct (d) auralization. See Fig.~\ref{fig:grids} for illustrations of the geometries. The figure in parentheses states the maximum SH order $N$ that can be extracted from the grid. Note that $L$ for the surface grids refers to the number of pairs of data points. The cubical grids have an edge length of \SI{0.14}{m} and the spherical grids have a diameter of \SI{0.14}{m}. }
\label{tab:grid_sizes}
\end{table}

We used the HRTFs from~\cite{Bernschutz:DAGA2013} to compute the stimuli. Audio examples of the conditions tested in the present experiment are available at~\cite{Auralization_audio_examples}.

19 subjects (of which twelve were men, six were women and one preferred not to tell) between 23 and 42 years of age (median: 28 years) with self-reported unimpaired hearing participated in the experiment. The subjects were acoustics researchers, students in the Sound and Vibration MSc program or attended a course on music technology, as well as one student without such background. The median duration of the experiment was \SI{19}{min} (minimum: \SI{10}{min}, maximum: \SI{32}{min}).

% -----------------------------------------------------------------------------------
% -----------------------------------------------------------------------------------
% -----------------------------------------------------------------------------------
\section{Results}\label{sec:results}

\begin{figure}[tb]
    \centering
    %                         left bottom right top
    \includegraphics[width=\columnwidth,trim=0in 0in 0in 0in, clip=true]{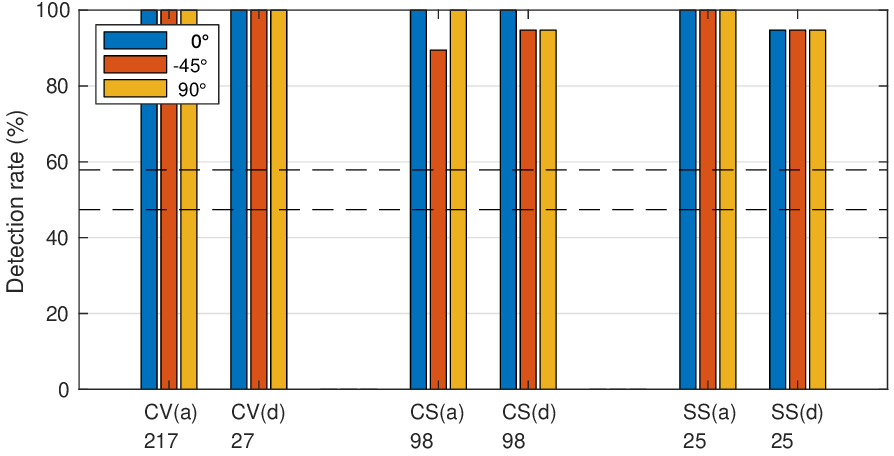}
    \caption{Obtained detection rates for the anechoic condition for low-density grids.  The dashed lines mark the maximum detection rate threshold for similarity (\SI{47.4}{\%}) and the minimum detection rate threshold for significant differences (\SI{57.9}{\%}).}
    \label{fig:detection_rates_1}
    %\vspace{-2mm}
\end{figure}
\begin{figure}[tb]
    \centering
    %                         left bottom right top
    \includegraphics[width=\columnwidth,trim=0in 0in 0in 0in, clip=true]{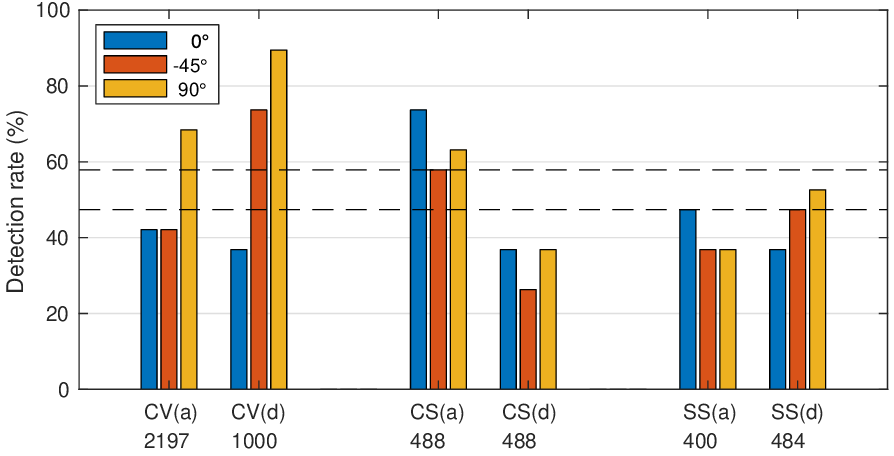}
    \caption{Obtained detection rates for high-density grids (anechoic condition). The dashed lines mark the maximum detection rate threshold for similarity (\SI{47.4}{\%}) and the minimum detection rate threshold for significant differences (\SI{57.9}{\%}).}
    \label{fig:detection_rates_2}
    %\vspace{-2mm}
\end{figure}
\begin{figure}[tb]
    \centering
    %                         left bottom right top
    \includegraphics[width=\columnwidth,trim=0in 0in 0in 0in, clip=true]{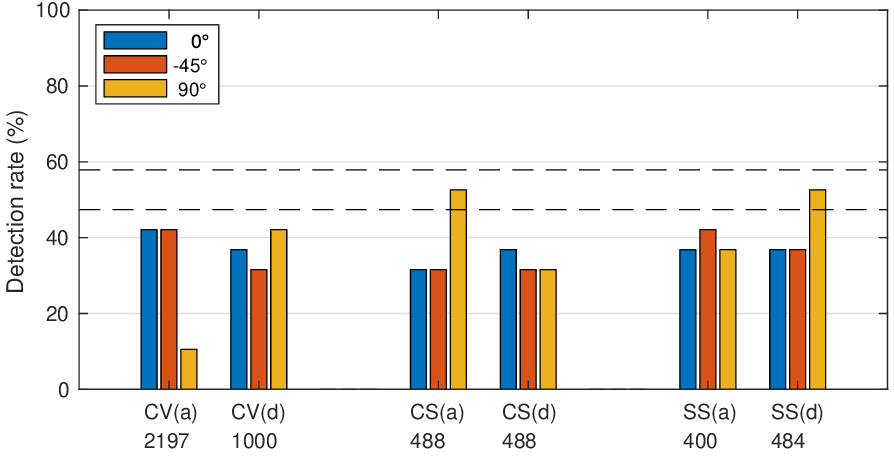}
    \caption{Obtained detection rates for high-density grids (reverberant condition). The dashed lines mark the maximum detection rate threshold for similarity (\SI{47.4}{\%}) and the minimum detection rate threshold for significant differences (\SI{57.9}{\%}).}
    \label{fig:detection_rates_3}
    %\vspace{-2mm}
\end{figure}

Fig.~\ref{fig:detection_rates_1}-\ref{fig:detection_rates_3} depict the obtained detection rates, i.e.~the percentage of trials in which the odd stimulus was correctly reported, for the different tested conditions. The standardized method to evaluate the results of a triangle test is based on a binomial model. If all subjects give random answers, i.e., if the auralization is perceptually transparent, the expected  detection rate is $33.\bar{3}~\%$. By defining an $\alpha$-level (the maximum acceptable probability of concluding that a perceptible difference exists when one does not), one can use the binomial distribution to calculate the minimum number of correct responses required to conclude that a perceivable difference exists. For the $N = 19$ subjects in this experiment, at least 11 correct detections (a detection rate of 57.9\%) are needed to reject the null hypothesis of perceptual similarity at $\alpha = 0.05$~\cite{ISO4120:2021}.

However, the absence of evidence for a perceivable difference does not imply evidence for the absence of a difference. A detection rate below the threshold does not automatically indicate perceptual transparency. To formally test for similarity, one can define a 
similarity threshold, $p_d$, which corresponds to the maximum allowable proportion of participants who are true ``discriminators'' (i.e., able to perceive a difference). This should not be confused with the raw detection rate, as non-discriminators can still guess correctly by chance. To account for this, one defines a $\beta$-level, limiting the probability of a Type II error (i.e., falsely concluding similarity when a perceptible difference exists), and uses the normal approximation to the binomial distribution to calculate an upper confidence limit for $p_d$, as described in~\cite{ISO4120:2021}.

In this study, we set $p_d = 0.5$ and $\beta = 0.05$, resulting in a threshold of up to 9 correct identifications (\SI{47.4}{\%}) for the stimuli to be considered sufficiently similar. This means that even if up to \SI{50}{\%} of the subjects can detect a difference, the stimuli are still considered perceptually similar within the chosen confidence level of \SI{95}{\%}. While this threshold may appear generous for claiming perceptual transparency, we argue that the triangle test represents a highly sensitive method for difference detection. First, such a direct comparison to the ground truth is rarely possible in applied settings. Second, any practical acoustic simulation method introduces systematic uncertainties (e.g., numerical dispersion), which are likely greater than the residual deviations observed under idealized auralization conditions with high-resolution grids and anechoic conditions.

The percentage of correctly identified differences between auralization and ground truth accumulated over all subjects are depicted in Fig.~\ref{fig:detection_rates_1}-\ref{fig:detection_rates_3}, with horizontal lines indicating both the previously discussed minimum threshold for statistically significant differences (\SI{57.9}{\%}) and the maximum threshold for sufficient similarity (\SI{47.4}{\%}). The results for low-density grids under anechoic conditions are shown in Fig.~\ref{fig:detection_rates_1} and confirm that participants were clearly able to identify differences independent of grid type, auralization method, and incidence angle.  

For high-density grids in anechoic environments, the perceptual results shown in Fig.~\ref{fig:detection_rates_2} are not as consistent. However, except for the \SI{90}{\degree} incidence direct condition, all detection rates for the spherical surface grid are below the similarity threshold, which means we assume perceptual transparency. While the direct auralization of the cubical surface grid appears to be indistinguishable from the ground truth as well, the detection rate for the ambisonic auralization of the same grid exceeds the threshold for significant differences independent of incidence angle. For the high-density volumetric cube in anechoic conditions, the incidence angle appears to have a greater impact on perceivable differences, as the detection rates for \SI{0}{\degree} auralizations are well below the similarity threshold, whereas the detection rates for \SI{90}{\degree} auralizations exceed the threshold for significant differences.

Finally, the results for the high-density grids in a reverberant sound field shown in Fig.~\ref{fig:detection_rates_3} confirm that, for most of the evaluated conditions, the auralizations can be considered as perceptually transparent compared to the ground truth. The only exceptions are the cubical surface grid auralized via ambisonic and the direct auralization of the spherical surface grid, which both slightly exceeded the similarity threshold for \SI{90}{\degree} incidence.

\section{Discussion}\label{sec:discussion}

It was reported in the literature that authentic binaural reproduction of reverberation can be performed at considerably lower SH orders than the direct sound~\cite{Engel:JASA2021,Lubeck:JASA2022}. This refers to cases were direct sound and reverberation were auralized separately. The targeted applications in the present case do not allow for separating direct sound and reverberation before auralization. The present experiment suggest that reverberation reduces perceived differences between auralization and ground truth somewhat. Our results from~\cite{Ahrens:DAGA2025} suggest that the properties of the direct sound appear to dominate the detectability of differences also in the reverberant case. Reverberation does not seem to have a large effect on the required signal processing for perceptual transparency. A final proof is to be presented.

The study that was present in~\cite{Lubeck:JASA2022} investigated the required SH order for perceptually transparent ambisonic auralization of recordings from rigid spherical microphone arrays, which is a setup that is very similar to the spherical surface grids in our study. The authors found that the median SH order required was approx.~20, which is in a similar order of magnitude like the SH order required for perceptual transparency in the present case. We conduced a variety of pilot studies that suggest that the results from~\cite{Lubeck:JASA2022} (and from~\cite{Zaunschirm:JASA2018,Ahrens:JASA2019,Engel:JASA2021} for that matter) are only partially comparable to the present ones. The reason for this is that these works used a reference signal for the comparison that was the output of a very-high-SH-order version of the rigid spherical microphone array that was being investigated. The consequence is that the reference and the stimuli under test differ only with respect to spatial aliasing. 

In the present study, the ground truth was not processed by the array pipeline. We found that this does make a noticeable difference when an experiment paradigm that is as sensitive as the triangular test is being used. We tested the signal processing from~\cite{Sheaffer:TASLP2015,Lubeck:JASA2022,BalmagesRafaely2007}in the present context and did not succeed in producing auralizations that sounded nearly as close to our non-array ground truth as when the present signal processing with the regularization based on the SVD is employed (recall Sec.~\ref{sec:methods}). A similarly regularized SVD was already identified in~\cite{Politis:WASPAA2017} to provide higher numerical accuracy with respect to a variety of metrics particularly at low frequencies. We confirm that it is preferable also in perceptual terms and provides benefits over the entire frequency range.

We found that the generally accepted assumption that arrays produce sufficiently accurate binaural output signals below the spatial aliasing frequency is not strictly applicable. We are confident to state that in the cases where our subjects were hearing differences for the high-density grids, the contribution of spatial aliasing to the difference was small. This is supported also by the numerical data shown in Fig.~\ref{fig:data6}: Condition CS(d) 488 is perceptually transparent, but the corruption of the signal in the contralateral ear due to spatial aliasing is considerable for the \SI{90}{\degree} incidence direction.

Our informal experimentation revealed that differences can even be apparent when the source signal is lowpass filtered such that no spatial aliasing is being triggered. This is the reason why we chose not to aim at identifying a minimum number of sampling points that guarantees perceptual transparency. Increasing the number of sampling points reduces spatial aliasing, but does not reduce the possible differences to the ground truth below the spatial aliasing frequency. We have not been able to identify these differences below the spatial aliasing frequency numerically, and we will direct our future research towards clarifying this aspect before identifying the minimum required number of sampling points. The found that the detectable differences below the spatial aliasing frequency occur over several octaves, which makes it unlikely that choosing different grid dimensions can mitigate this.

It appears that the exact choice of the regularization that needs to be applied with any of the tested methods is decisive. A substantial amount of trial-and-error led to the choice of frequency-dependent regularization parameters that we used in the preparation of the stimuli. It is conceivable that  regularization parameters can be found for all tested high-density grids to produce perceptually transparent auralization in the strict sense.

We need to remind the reader that the above discussion assumes comparing auralization and ground truth in a triangular test. Other settings like casual listening or potentially even A-B comparisons (without a third alternative) may not reveal the differences that are discussed above. We encourage the reader to experience this for themselves based on the audio examples that we provide in~\cite{Auralization_audio_examples}.

The work presented in~\cite{Cosnefroy:I3DA2023} also used virtual cardioid sensors for ambisonic auralization of sampled sound fields and confirmed high numerical accuracy of the binaural signals for an SH order of~16. Note that the authors used slightly different signal processing than us, and only data up to a frequency of \SI{10}{kHz} are shown. Perceptual data are not provided.

%\color{black}

% -----------------------------------------------------------------------------------
% -----------------------------------------------------------------------------------
% -----------------------------------------------------------------------------------
\section{Conclusions}

Our work showed that the perceptually transparent auralization over the entire audible frequency range was possible with some of the tested methods and parameter choices. This is an important result as it ensures that data from any numerical simulation methods that can compute sound pressure and particle velocity at the required sampling points can be auralized faithfully. The validation paradigm employed was very sensitive so that even parameter sets that did not provide perceptual transparency in the strict sense are likely to be very useful in practice given the typical numerical uncertainties and bandwidth limitations of simulation methods.

We can confirm through extensive experimentation that details of the implementation of the signal processing can be decisive for whether perceptual transparency can be achieved or not. We provide a free implementation of all investigated methods in~\cite{Ahrens:auralization_toolbox} for reproducibility. It is not inconceivable that further improvements of the signal processing can allow for even more accurate auralization. \revised{Specifically, we found that audible differences can remain below the spatial aliasing frequency even when carefully selected signal processing parameters are employed. Overcoming this is subject to future research.}

It may not be convenient in many situations to design the sampling grid on which a given acoustic simulation is computed such that the simulation grid is most favorable for auralization. Interpolation needs to be applied in these cases. The perceptual implications of this are to be investigated.

The investigated methods have the potential of serving as a universal auralization methods as they can be applied to any acoustic simulation data so long as the data can be converted to a sound pressure and/or particle velocity distribution. The simulation data can originate from a variety of source including geometric acoustics simulations, wave-based simulations, and radiocity. This can make simulation data more straightforward to compare and allows for broadening the scope of round-robin tests like~\cite{Brinkmann:JASA2019}. The code base that accompanies this article provides example projects for auralization of data from commercial room acoustic simulation softwares with the presented methods.

% -----------------------------------------------------------------------------------
% -----------------------------------------------------------------------------------
% -----------------------------------------------------------------------------------
\section{Resources}

Implementations of the methods that we investigated in this article are available freely in the Chalmers Auralization Toolbox~\cite{Ahrens:auralization_toolbox}. Audio examples are available at~\cite{Auralization_audio_examples}.

%%\bibliographystyle{abbrv}
%\bibliographystyle{jaes}
%\interlinepenalty=10000
%\bibliography{references}
%\interlinepenalty=0

\break

%Appendix
%\appendix

%\section*{APPENDIX}

%Appendix goes here

\end{document}